\documentclass[aip,amsmath,amssymb,reprint]{revtex4-1}
\pdfoutput=1
\usepackage{graphicx}
\usepackage{dcolumn}
\usepackage{natbib}
\usepackage{bm}

\usepackage[utf8]{inputenc}
\usepackage[T1]{fontenc}
\usepackage{mathptmx}
\usepackage{etoolbox}

\usepackage{algorithm2e}
\RestyleAlgo{ruled}
\SetKwComment{Comment}{// }{} 

\SetCommentSty{commentfont} 
\DontPrintSemicolon
\SetKwProg{Fn}{Function}{:}{}

\usepackage{color}
\usepackage{listings}
\usepackage{xcolor}

\definecolor{codegreen}{rgb}{0,0.6,0}
\definecolor{codegray}{rgb}{0.5,0.5,0.5}
\definecolor{codepurple}{rgb}{0.58,0,0.82}
\definecolor{backcolour}{rgb}{0.95,0.95,0.92}

\lstdefinestyle{mystyle}{
    backgroundcolor=\color{backcolour},   
    commentstyle=\color{codegreen},
    keywordstyle=\color{magenta},
    numberstyle=\tiny\color{codegray},
    stringstyle=\color{codepurple},
    basicstyle=\ttfamily\footnotesize,
    breakatwhitespace=false,         
    breaklines=true,                 
    captionpos=b,                    
    keepspaces=true,                 
    numbers=left,                    
    numbersep=5pt,                  
    showspaces=false,                
    showstringspaces=false,
    showtabs=false,                  
    tabsize=2
}
\makeatletter
\def\@email#1#2{%
 \endgroup
 \patchcmd{\titleblock@produce}
  {\frontmatter@RRAPformat}
  {\frontmatter@RRAPformat{\produce@RRAP{*#1\href{mailto:#2}{#2}}}\frontmatter@RRAPformat}
  {}{}
}%
\makeatother
\begin{document}

\preprint{AIP/123-QED}

\title[mmodel]{mmodel: A workflow framework to accelerate the development of experimental simulations}
\author{Peter Sun}
\email{hs859@cornell.edu}
\author{John A. Marohn}%

\affiliation{ 
Department of Chemistry and Chemical Biology, Cornell University}

\date{\today}

\begin{abstract}
Simulation has become an essential component of designing and developing scientific experiments. 
The conventional procedural approach to coding simulations of complex experiments is often error-prone, hard to interpret, and inflexible, making it hard to incorporate changes such as algorithm updates, experimental protocol modifications, and looping over experimental parameters.
We present \textit{mmodel}, a framework designed to accelerate the writing of experimental simulation packages.
\textit{mmodel} uses a graph-theory approach to represent the experiment steps and can rewrite its own code to implement modifications, such as adding a loop to vary simulation parameters systematically.
The framework aims to avoid duplication of effort, increase code readability and testability, and decrease development time.

\end{abstract}

\maketitle

\section{Introduction} 
Scientific experiments are becoming ever more complex, with multiple interconnected components.
Experiments thus often require simulation code to explore the feasibility, optimize experimental parameters, and extract information from experimental results.
In fields such as mass spectrometry,
\citep{Snyder2016jan,Nolting2019mar} X-ray spectroscopy,\citep{Fadley2010may,Gann2012apr,Norman2018aug} and 
magnetic resonance force microscopy (MRFM),\citep{Rugar2004jul, Degen2009feb, Moore2009dec, Vinante2011dec, Nichol2012feb} experiments are expensive and time-consuming.
In these and other fields, simulation has become crucial to experimental design and hardware development.
For example, research groups in the MRFM field employ a wide variety of experimental setups, materials, and spin-detection protocols.
In order to design experiments and validate results across laboratories, we need simulation codes that cover all possible setups, materials, and protocols.
The code should be fast, readable, reliable, modular, and expandable.

The typical approach to experimental simulation is to create one-off simulations for each experiment of interest.\citep{Trisovic2022feb}
Unfortunately,  the code needs to be rewritten when we modify the experiment or implement a loop to vary parameters efficiently.
These practices result in simulations that contain large amounts of duplicated code and are often unreadable.
Moreover, the one-off nature of the code impedes reproducibility and leads to errors due to a lack of testing and annotation;\citep{Trisovic2022feb, Pimentel2019may} coding errors lead to incorrect publication results in many scientific fields.\citep{Bishop2018sep, Perkel2022feb, Strand2023jan}
A new approach is needed to develop modular, expandable, testable, and readable simulation code.

This paper introduces the \textit{mmodel} package that provides a framework for coding experimental simulations. 
\textit{mmodel} employs a graph-based representation of the experimental components, a method popular in scientific workflows design.\citep{Deelman2009may,Liew2017dec}
The framework prioritizes the prototyping phase of the development by allowing modification to the simulation model post-definition, which most workflow systems cannot achieve.
In addition, we show that existing projects can easily migrate to the \textit{mmodel} framework.  
Section~\ref{example} introduces a simple representative physics experiment and shows how \textit{mmodel}'s graph-theory-based simulation approach facilitates experimental parameter exploration, experimental result validation, and the debugging process. 
Section~\ref{discussion} compares \textit{mmodel} to other popular scientific workflows and discusses the advantages of \textit{mmodel} in the prototyping phase of development.
Appendix \ref{API} summarizes the \textit{mmodel} API and presents example scripts to define, modify and execute the experiment discussed in Section \ref{example}.

\section{\label{example}Scientific experiment prototype} 
\begin{figure}[t]
\centering
\includegraphics[width=\columnwidth]{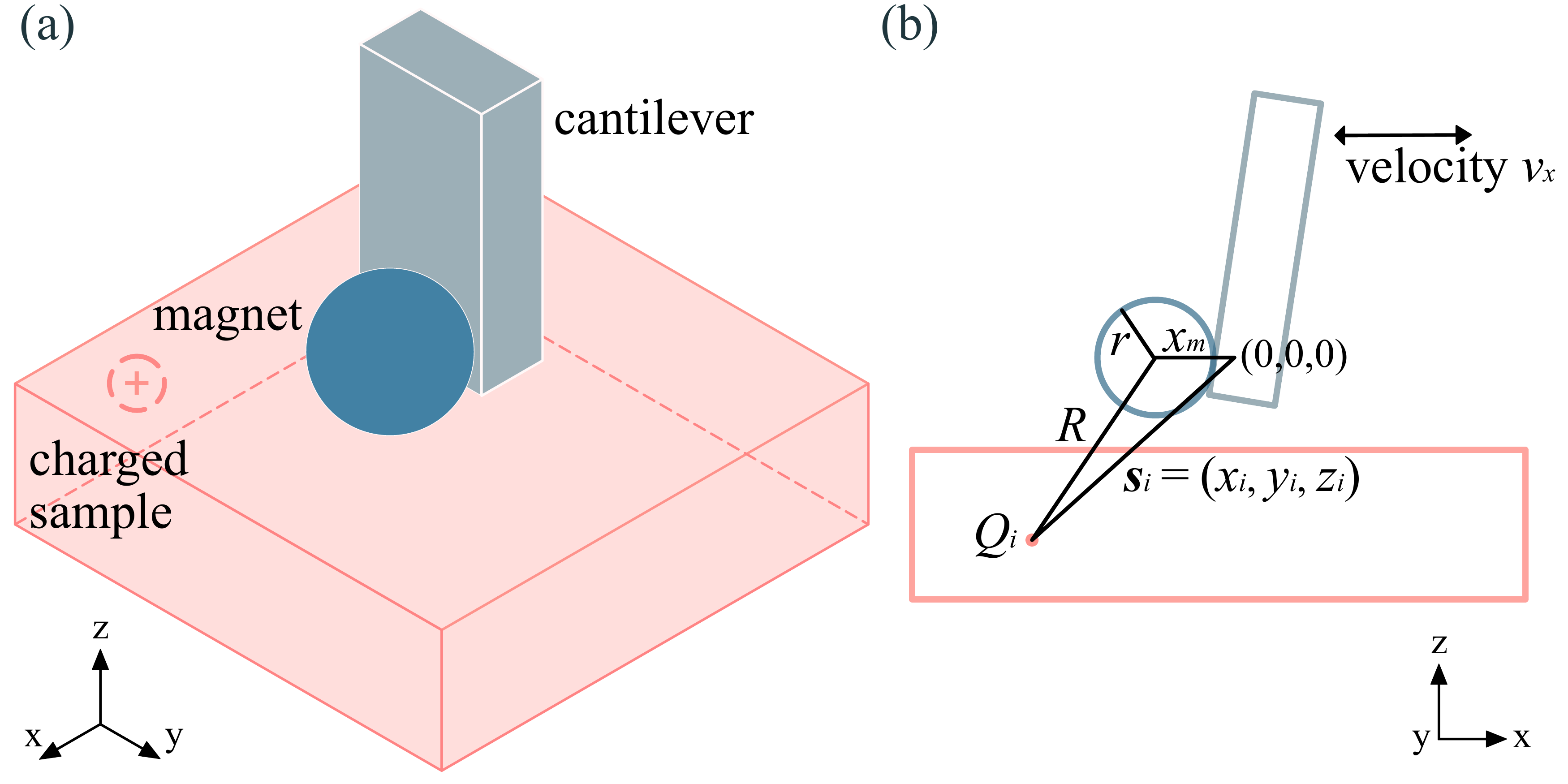}
\caption{
Example setup. 
(a) The isometric view. 
A magnet is attached to a micro-cantilever.
A charged sample is located below the magnet.
(b) The side view.
The cantilever oscillates in the $x$ direction, and the magnet is $x_\mathrm{m}$ away from the equilibrium position. The $i^\mathrm{th}$ sample charge $Q_i$ is located at $\bm{s}_i = (x_i, y_i, z_i)$.
}
\label{fig:setup}
\end{figure}

Consider the following contrived experiment investigating the force exerted on a moving magnet by a charged sample.
The magnet is spherical and is attached to a micro-cantilever that oscillates in the $x$ direction, shown in Fig.~\ref{fig:setup}.
The force on a moving charge $q$ by an electric field $\bm{E}$ and a magnetic field $\bm{B}$ can be calculated using the Lorentz force law \citep{Griffiths2013}
\begin{equation}
\bm{F} = q(\bm{E} + \bm{v} \times \bm{B}).
\end{equation}
Using the magnet as the frame of reference, the problem is equivalent to calculating the sum of the force exerted on a fixed magnet by moving charges traveling through the magnet's magnetic field.\citep{Jefimenko1993mar}

In our contrived experiment, the velocity has only an $x$ component, $v_x$.
We can approximate the cantilever oscillation as simple harmonic motion in the $x$ direction.
Given the cantilever frequency $f$ and amplitude $A$, the magnet displacement $x_{\mathrm{m}}$ (measured relative to the equilibrium position) and the velocity $v_x$ are
\begin{equation}
x_{\mathrm{m}}(t) = A\cos(\omega t)
\end{equation}
\begin{equation}
v_x(t) = -\omega A \sin(\omega t)
\end{equation}
where $\omega = 2\pi f$ and $x_{\mathrm{m}} \in (-A, A)$.
We assume the magnet's origin at the equilibrium position is $(0, 0, 0)$.
We want to calculate the force on the magnet when the displacement is $x_{\mathrm{m}}$.
The instantaneous magnet velocity $v_x$ at the position of motion ($x_{\mathrm{m}}$, 0, 0) can be calculated as follows:
\begin{equation}
v_x(x_{\mathrm{m}}) = \pm 2 \pi f \sqrt{A^2 - x_{\mathrm{m}}^2}.
\end{equation}

The sphere's magnetic field in the $y$ and $z$ directions at the $i^{\mathrm{th}}$ charge's position $\bm{s_i} = (x_i, y_i, z_i)$, $B_y(\bm{s}_i)$ and $B_z(\bm{s}_i)$, can be calculated using
\begin{equation}
B_z (\bm{s}_i) = \frac{\mu_0 M_{\mathrm{s}}}{3}\left( 3 \frac{Z_i^2}{R_i^5} - \frac{1}{R_i^3} \right)
\end{equation}
\begin{equation}
B_y (\bm{s}_i) = \mu_0 M_{\mathrm{s}} \left( \frac{Y_i Z_i}{R_i^5}\right),
\end{equation}
with $\mu_0$ the vacuum permeability;
$M_{\mathrm{s}}$ the magnet saturation magnetization;
$r$ the magnet radius;
$X_i = (x_i-x_{\mathrm{m}})/r$, $Y_i = y_i/r$, and $Z_i = z_i/r$ the reduced charge coordinates; and
$R_i = \sqrt{X_i^2+Y_i^2+Z_i^2}$ a reduced distance.
It is important to note that each point charge can induce an image charge in the magnet, resulting in an additional electric and magnetic field.\citep{Griffiths2013}
We ignore the fields generated based on this effect for simplicity.

Finally, without the external field $\bm{E}$, the forces exerted on the magnet in the $y$ and $z$ direction, $F_y$ and $F_z$, by a set of $N$ discrete charges $Q_i$ can be calculated by summing the forces exerted by the individual charges:
\begin{equation}
F_y(x_{\mathrm{m}}) = v_x(x_{\mathrm{m}}) \sum_{i=1}^{N} Q_i B_z(\bm{s}_i, x_{\mathrm{m}})
\end{equation}
\begin{equation}
F_z(x_{\mathrm{m}}) = - v_x(x_{\mathrm{m}}) \sum_{i=1}^{N} Q_i B_y(\bm{s}_i, x_{\mathrm{m}}),
\end{equation}
where $Q_i = Q(\bm{s}_i)$ is the charge of the $i^{\mathrm{th}}$ sample charge located at $\bm{s}_i$.
In the pseudocode, we use $grid$ to represent the collection of charge positions.
The variables are shown in Table \ref{table:1}.

\begin{table}[t]
\caption{\label{table:1}List of the Lorentz force law model variables calculating the forces on moving magnet from a charge distribution.}
\begin{ruledtabular}
\begin{tabular}{ll}

 variable & description\\ 
 \hline
$f$ & cantilever frequency  \\ 
$A$ & cantilever amplitude  \\ 
$v_x$ & instantaneous magnet motion velocity  \\ 
$x_{\mathrm{m}}$ & magnet position in the $x$ direction\\
$r$ & magnet radius\\ 
$\mu_0$ & vacuum permeability \\ 
$M_{\mathrm{s}}$ & magnet saturation magnetization \\ 
$B_y$ & magnet magnetic field in the $y$ direction \\ 
$B_z$ &magnet magnetic field in the $z$ direction \\ 
$q$ & point charge \\ 
$Q_i$ & charge of the $i^{\mathrm{th}}$ sample charge \\ 
$\bm{s}_i$ & location of the $i^{\mathrm{th}}$ sample charge, $(x_i, y_i, z_i)$ \\ 
$grid$ & collection of sample charge positions \\
$F_y$ & force on the magnet  in the $y$ direction\\
$F_z$ & force on the magnet in the $z$ direction \\

\end{tabular}
\end{ruledtabular}
\end{table}

\subsection{Procedural Approach}

Conventionally, simulations use a procedural approach that executes the experiment in steps.
One procedural approach to calculating the force on the point charge is to group the calculation by component.
We use \texttt{Bzfunc}, \texttt{Byfunc}, \texttt{Vxfunc}, \texttt{Fyfunc}, and \texttt{Fzfunc} to denote the respective calculations. 
Here we first calculate the components $v_x$, $B_x$, and $B_y$. 
The second step is to calculate $F_y$ and $F_z$ based on the values of $Q$, $v_x$, $B_z$, $B_y$. 

In the calculation, we step through the grid points to calculate the effective force at each point of the charge distribution, and we sum all the force values to obtain the total force acting on the cantilever.
The algorithm is shown in Alg.~\ref{alg:1}.
In the prototyping process, the above calculations can be packaged into a function or a script, with the inputs of the magnet, cantilever, and grid as input parameters and the $y$- and $z$-direction forces as outputs. 

\begin{algorithm}[t]
\SetKwFunction{Func}{Experiment}
\Fn{\Func{$A, f, grid, \mu_0, M_{\mathrm{s}}, Q, r, x_{\mathrm{m}}$}}{
    $v_x \gets \mathrm{Vxfunc}(f, A, x_{\mathrm{m}})$\;
    $B_z \gets \mathrm{Bzfunc} (\mu_0, M_{\mathrm{s}}, r, grid, x_{\mathrm{m}})$\;
    $B_y \gets \mathrm{Byfunc} (\mu_0, M_{\mathrm{s}}, r, grid, x_{\mathrm{m}})$\;
    \For{$x, y, z$ in $grid$}{
    $F_{y, grid}[x, y, z] \gets Q[x, y, z] \times v_x \times B_z[x, y, z]$\;
    $F_{z, grid}[x, y, z] \gets - Q[x, y, z] \times v_x \times B_y[x, y, z]$\;
    }
$F_y \gets sum(F_{y, grid})$\;
$F_z \gets sum(F_{z, grid})$\;
}{\KwRet{$F_y,F_z$}}
\caption{Procedural program of the Lorentz force law model calculating the forces on moving magnet from a charge distribution.}
\label{alg:1}
\end{algorithm}

\subsection{\textit{mmodel} graph-theory approach}

Alternatively, we can represent the process flow in a directed acyclic graph (DAG) $G = (V, E)$, with nodes $V$ as the experimental steps and edges $E$ as the data flow.
A DAG is a directed graph with no directed cycles. A DAG has at least a topological ordering, where for all the edges $(u, v)$, the parent node $u$ is always positioned before the child node $v$. 
With this approach, we create a graph representing the experiment and simulation process and execute the nodes in topological order while managing the intermediate value flows.
In \textit{mmodel}, each process is a Python callable.
 
\begin{figure}[t]

\centering
\includegraphics[width=\columnwidth]{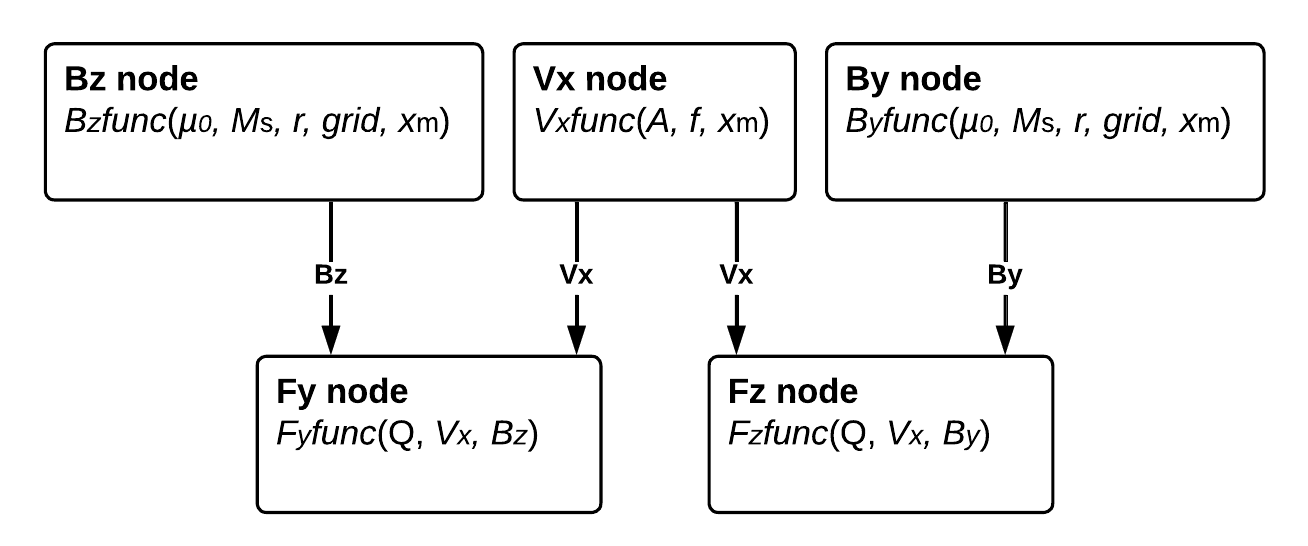}
\caption{ \textit{mmodel} graph-theory representation of the Lorentz force law model calculating the forces on a moving magnet from a charge distribution.}
\label{fig:bd-full}
\end{figure}

Here we divide the experiment into five nodes, named by their output: Vx, Bz, By, Fy, and Fz.
Each node corresponds to its respective functions, resulting in a graph $G = \{V, E\}$, where $V = \{\mathrm{Vx, Bz, By, Fy, Fz}\}$ and $E = \{\mathrm{(Bz, Fy), (By, Fz), (Vx, Fy), (Vx, Fz)}\}$, as shown in Fig.~\ref{fig:bd-full}.
Each node takes input parameters externally or from its parent node. The graph is then converted into a function by defining the method of execution of the notes in the topological ordering.
The ordering ensures that the parent node of an edge is always executed first.

One topological ordering for graph G is [Bz, By, Vx, Fy, Fz]. 
To direct the data flow, we can store the intermediate value of $v_x$, $B_z$, and $B_y$ calculated by the corresponding nodes.
These intermediate values can be stored internally in memory or externally on disk.
The function returns the output of the two terminal nodes, Fy and Fz. 

\subsection{Experimental exploration: Investigate the effect of different cantilever frequencies} \label{case:ee}

To investigate the effect of cantilever frequency $f$ over the forces on the charge, we want to loop through $f$ values and examine the resulting changes in the force acting on the cantilever.
Since calculating $B_y$ and $B_z$ does not depend on the $f$ parameter, the most efficient simulation is to loop only the \texttt{Vxfunc}, \texttt{Fyfunc}, and \texttt{Fzfunc} calculations. 
For the procedural approach, a new function must be written by modifying the original experiment source code to add loops to the \texttt{Vxfunc} subsection of the code.
 
\begin{figure}[t]

\centering
\includegraphics[width=0.8\columnwidth]{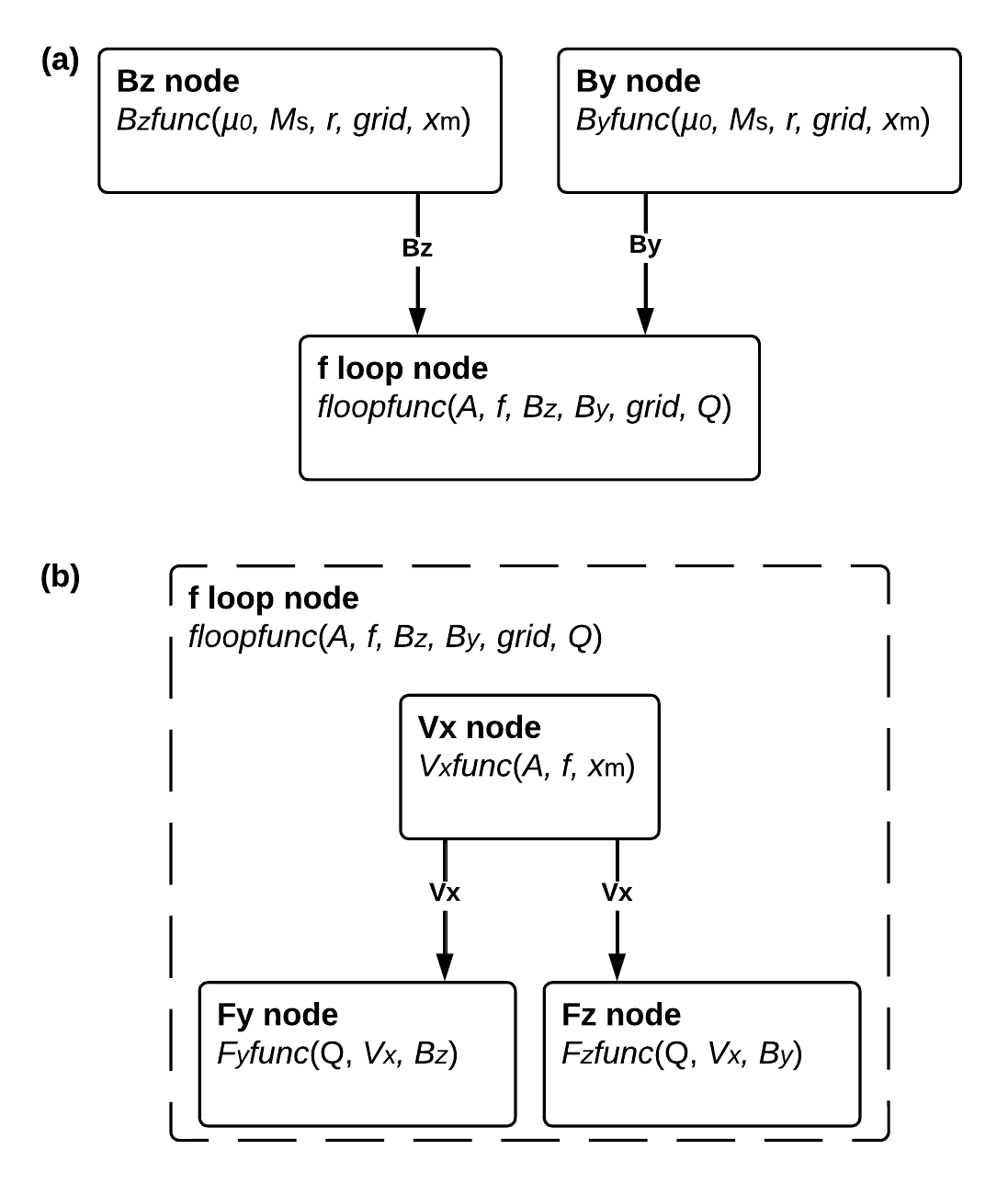}
\caption{\textit{mmodel} graph-theory representation of the Lorentz force law model with the frequency $f$ loop node.
(a) The new mmodel graph with subgraph substituted with a loop node, (b) $f$ loop node model graph.}
\label{fig:bd-loop}
\end{figure}

However, the graph-theory approach can achieve this looping result without modifying the original code. 
Adding the loop to the workflow can be achieved in four steps.
First, we create a subgraph, $H =  \{V = \{\mathrm{Vx, Fy, Fz}\}, E = \{\mathrm{(Vx, Fy), (Vx, Fz)}\}\}$, that excludes the By and Bz nodes from the original graph.  
Second, we convert the subgraph into a function with the additional loop step. 
Third, we replace the subgraph with a single node in the original graph and map it to the subgraph function. 
The new graph is shown in Fig.~\ref{fig:bd-loop}, where \texttt{floopfunc} is the function that loops through the $f$ values.
Last, the newly obtained graph is converted into a new experimental function.
The process can be combined into a single function, shown in Alg.~\ref{alg:loop}, and applies to all graphs defined using \textit{mmodel}'s graph definition.
Importantly, these steps can be automated, requiring only a one-line code modification to implement parameter looping (see \ref{subsec:looping}).

\begin{algorithm}[t]
\caption{Function to create a looped subgraph within a given graph using the \textit{mmodel} graph-theory approach.}\label{alg:loop}
\SetKwFunction{Func}{LoopMod}
\Fn{\Func{\rm{G, loopParam}}}{\Comment{G is the original graph}
    H $\gets$ subgraph(G, loopParam)\;
    \Comment{nodes with loopParam as input parameter and their child nodes}
    loopFunc $\gets$ loopmodel(H, loopParam)\;
    \Comment{convert the graph into a function with added loop}
    loopG $\gets$ replace(G, H, loopFunc)\;
    \Comment{replace the subgraph H in G with a new node and map loopFunc to it}
}
{\KwRet{\rm{loopG}}}
\end{algorithm}

\subsection{Experimental validation: Determine the cantilever amplitude}\label{case:ev}

In an experiment, the cantilever is driven at its resonance frequency, and the cantilever amplitude $A$ can be measured by the position detector.
Suppose we have measured $F_y$ and $F_z$ and want to use these values to validate the amplitude value.
In a procedural approach, a function is written to iterate through samplings (i.e., Monte Carlo samplings) of $A$ values until the simulated forces agree with the experiment (within a specified error).
Similar to Section \ref{case:ee}, only the \texttt{Vxfunc}, \texttt{Fyfunc}, and \texttt{Fzfunc} depend on $A$.

With the graph-theory approach, we can create a function that iterates through given parameter values until the termination condition is met. 
The function \texttt{SampleMod} that creates a new graph with the Monte Carlo fitting process is shown in Alg.~\ref{alg:ev}.
For the example scenario, the subgraph containing the Vx, Fy, and Fz nodes is converted into a function.
The \texttt{MonteCarlo} function takes the subgraph function as an input and outputs the optimal offset value.

\begin{algorithm}[t]
\caption{Function to add Monte Carlo fitting step to a given graph using the \textit{mmodel} graph-theory approach.}\label{alg:ev}
\SetKwFunction{Func}{SampleMod}
\Fn{\Func{\rm{G, iterParam, condition}}}{
\Comment{G is the original graph}
H $\gets$ subgraph(graph, iterParam)\;
\Comment{nodes with iterParam as input parameter and their child nodes}
subgraphFunc $\gets$ model(H, iterParam)\;
\Comment{convert the graph into a function}
sampleFunc  $\gets$ MonteCarlo(subgraphFunc, result, iterParam, condition)\; \Comment{iterate through values of iterParam against the experimental result with Monte Carlo sampling and terminate based on the condition}
sampleG $\gets$ replace(G, H, sampleFunc)\;
\Comment{replace the subgraph H in G with a new node and map sampleFunc to it}
}{\KwRet{\rm{sampleG}}}

\end{algorithm}

\subsection{Debugging: Investigate intermediate values} \label{case:debug}

In a prototyping process, debugging is crucial; common issues include data value or format errors.  
Sometimes we want to inspect the intermediate results, such as the $v_x$ or $F_y$ values.
Conventionally, this inspection is done by adding print or logging statements --- a significant effort.

\begin{figure}[t]

\centering
\includegraphics[width=0.6\columnwidth]{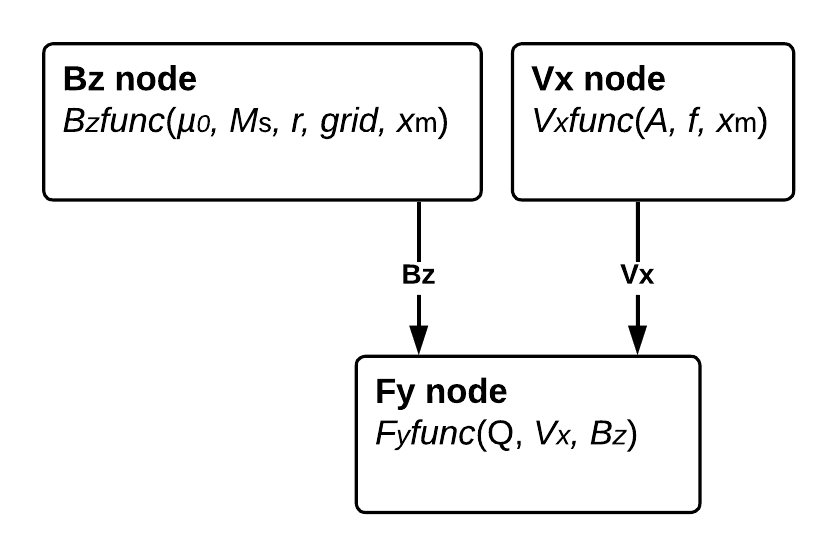}
\caption{Subgraph that only outputs $F_y$ value based on the Lorentz force law model graph.}

\label{fig:bd-partial}
\end{figure}

\begin{figure}[t]

\centering
\includegraphics[width=1.02\columnwidth]{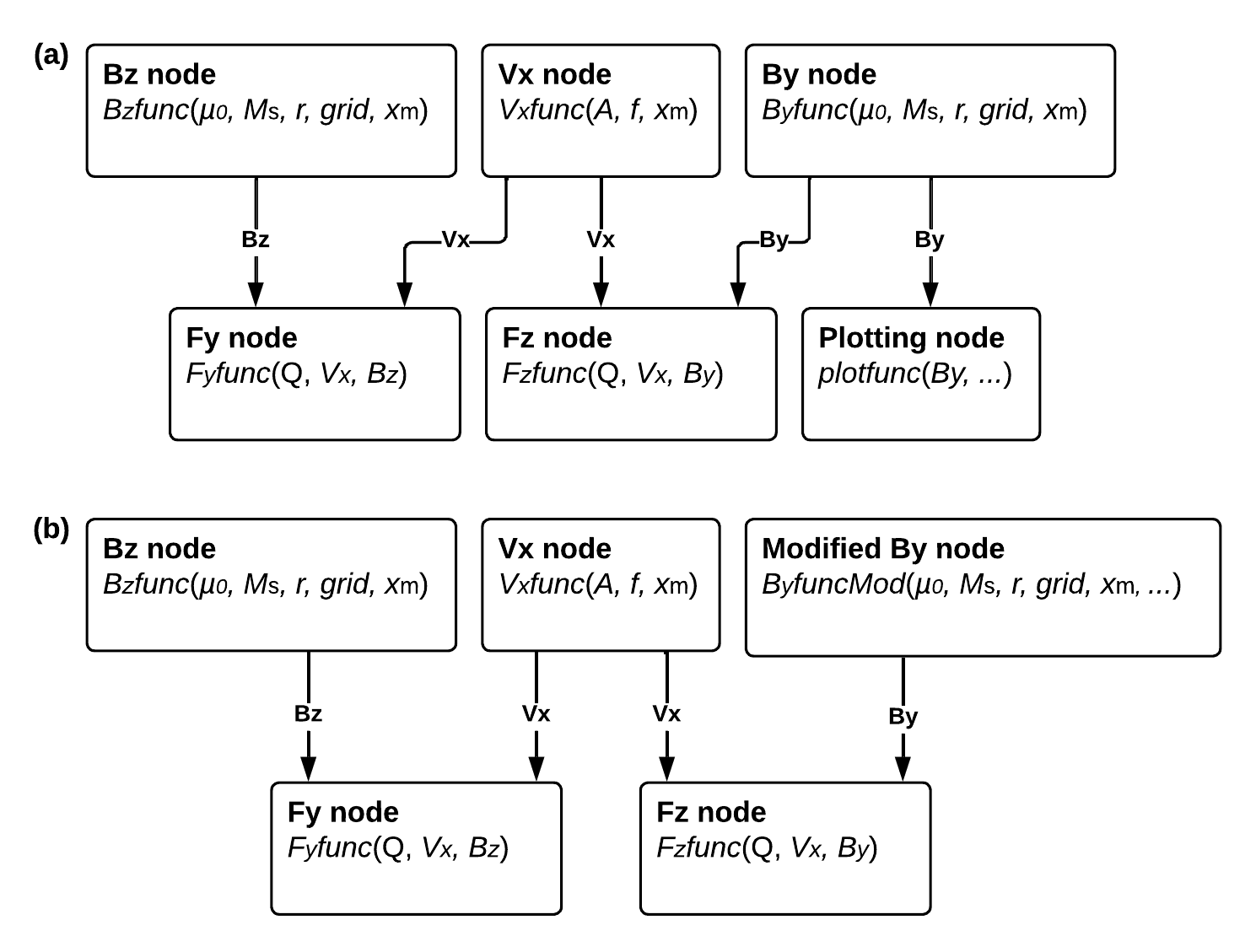}
\caption{Graphs with added plotting functionality based on the Lorentz force law model. (a) Add a node that plots By value to the graph, (b) modify the By node to add plotting functionality with calculation.}

\label{fig:bd-mod}
\end{figure}

Debugging becomes more straightforward with graphs: we can obtain the subgraph that outputs the desired intermediate results. 
The subgraph excludes unnecessary calculations, which makes it very efficient computationally. 
For example, we can obtain the subgraph $H = \{V = \{\mathrm{Bz, Vx, Fy}\}, E = \{\mathrm{(Bz, Fy), (Vx, Fy)}\}\}$ that only includes the Fy node and its parents, shown in Fig.~\ref{fig:bd-partial}. 
Additionally, since the process controls the output of the nodes, we can output the intermediate value alongside the original output. 
Alternatively, we can modify and add nodes to the graph to achieve additional functionality, such as plotting, shown in Fig.~\ref{fig:bd-mod}.

\section{\label{discussion}Discussion} 
The last two decades have seen an explosion of data in science and the data-driven science that comes with it. 
Scientific workflows become useful tools for managing and analyzing data, freeing domain experts from much tedious programming.\citep{Deelman2009may, Liew2017dec}
There are various popular packages and software that facilitate the simulation of scientific experiments, such as \textit{Pegasus},\citep{Deelman2019jul} \textit{Kepler},\citep{Ludascher2006aug, McPhillips2009may} \textit{AiiDA},\citep{Huber2020dec} \textit{Dask},\citep{Rocklin2015} and \textit{Pydra}.\citep{Jarecka2020}
These packages operate at different levels of complexity.
For example, \textit{Pegasus}, \textit{Kepler}, and \textit{AiiDA} focus on complete data pipeline management, while \textit{Dask} and \textit{Pydra} focus on data analysis.
However, these packages are most suited and optimized for upscaling scientific analysis with well-defined models. 
In our MRFM experiments, we constantly update the signal calculation algorithm and continuously loop over parameters and protocols to decide among experimental protocols and determine the optimal experimental conditions.
With the growing number of changes to the simulation code, neither the procedural approach nor the existing scientific workflow package suits the development process. 
The difficulty in maintaining the simulation package inspired the creation of \textit{mmodel}, which aims to reduce code duplication and provide automated solutions to simulation updates.

\begin{figure}[t]

\centering
\includegraphics[width=0.48\textwidth]{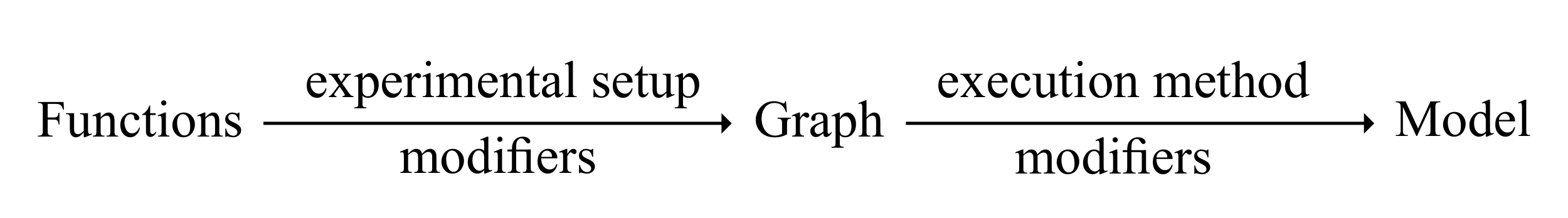}
\caption{\textit{mmodel} workflow definition steps. Modifiers can be applied to nodes and models. The graph defines the experimental steps, and the model defines the execution method.}
\label{fig:flow}
\end{figure}

As highlighted in the Section \ref{example}, \
\textit{mmodel} solves four issues when coding experimental simulations. 
First, the framework reduces code duplication and makes it easy to update subsections of the simulations.
\textit{mmodel} separates the code definition into three parts: function, graph, and model, shown in Fig.~\ref{fig:flow}. 
The functions describe the theoretical calculation of the experimental steps. 
The graphs describe the steps of the experiments, and the model defines the execution process. 
This separation makes it easy to only change or update one part of the process without modifying the other two. 
For example, in Section~\ref{example}, if we wanted to update the calculation of $v_x$, only the \texttt{VxFunc} function would need to be updated. 
In the prototyping phases, these changes often happen due to theory and algorithm updates and bug fixes. 
The code reduction resulting from using \textit{mmodel} requires fewer tests, which speeds up the development process.
Second, \textit{mmodel} simplifies experiment parameters testing. The graph in \textit{mmodel} is interactive, and all node objects are Python functions. There is no additional decorator required during the function definition step. This design allows defining graphs with existing functions, such as Python built-in or \textit{Numpy}-package functions. In addition, using functions as node objects extends the capability of the execution process. For example, a special function can be defined to identify subgraphs and create loops for a specific parameter. Our function node implementation extends the traditional capability of the DAG, which cannot contain cycles. We can write special functions to modify a process or a group of processes, such as the \texttt{LoopMod} function in Alg.~\ref{alg:loop}, which can loop a given parameter for all the graphs defined using \textit{mmodel}. Once these special functions are written, users can apply them directly to their graphs. 
Third, \textit{mmodel} optimizes runtime performance during the prototyping process. 
For example, \textit{mmodel} provides an execution method that optimizes memory---\texttt{mmodel.MemHandler}---by keeping a reference count of the times intermediate parameters are needed by the child nodes.
Once all child nodes have accessed the value, the value is deleted.
Another build-in execution method---\texttt{mmodel.H5Handler}---optimizes runtime memory by storing intermediate values in an H5DF file.
Additionally, debugging using the partial graphs or outputs of the intermediate value speeds up the prototyping cycle.
Fourth, \textit{mmodel} improves the readability of the code by providing a visual graph representation of the experiment with rich metadata.

\textit{mmodel} is advantageous over the other workflow systems in the prototyping phase. The existing scientific workflow packages are designed for well-defined models, making the computing steps definition well-structured.
As a result, the API is restrictive for modifying existing workflows and defining processes outside the package's scope.
Prototyping these packages creates unnecessarily complicated code and has a high learning curve.
For example, \textit{Pegasus} and \textit{Kepler} require developers to define the functions and graphs using a GUI with a domain-specific language.
\textit{AiiDA}, \textit{Dask}, and \textit{Pydra} require function definition with package-specific Python decorators and special node classes.
The function and graph definition cannot be easily transferred to another package due to their domain-specific approach.\citep{Liew2017dec}
These packages provide a limited ability to modify the graph post definition. For example, to add loop modification, they must either be written into the graph using a special node that directs the parameter inputs (\textit{Pegasus}, \textit{Pydra}) or built into the model (\textit{Kepler}, \textit{AiiDA}, \textit{Dask}). They cannot be added to the already-defined graph.
This implementation prevents post-interaction at the graph or model level. Any modification to the simulation process requires a completely new workflow.
In contrast, \textit{mmodel} is designed to be lightweight and flexible. Each part of the graph and model definition process allows for customization. For example, a new execution process can be written with the \texttt{mmodel.handler} API. The implementation also allows existing packages to use the framework easily with minimal adjustments to existing functions. The drawback of the \textit{mmodel}'s function-first approach is that sub-operation modification creates nested graphs and operations, making it difficult to upscale if many modification steps are required.
However, we believe the resulting modest increase in computational complexity is a reasonable compromise at the prototyping stage.

We believe \textit{mmodel} fills the void of a fast, reproducible, and general-purpose scientific workflow framework for simulating complex experiments.
\textit{mmodel} is written in Python with minimal API restrictions, making the framework excellent for building on or migrating existing Python codes.
The graph representation of the experiment and the associated rich metadata helps with communication among collaborators.
The flexibility and extendability allow code optimization and easy unit testing, reducing errors and accelerating code development to simulate complicated experiments.

\section{Future Directions} 
In future versions, we would like to provide more graph execution methods that optimize performance and profilers that provide execution analysis.
Additionally, we would like to add adaptors that can convert \textit{mmodel} graphs to other popular workflow frameworks.  
We encourage scientists and developers to contribute to the project. 
The \textit{mmodel} version 0.6 is available at the repository: \url{https://github.com/marohn-group/mmodel} and the documentation is available at \url{https://marohn-group.github.io/mmodel-docs/}.
The project has 100\% unit test coverage and is open-source under the 3-Clause BSD License.

\section{Acknowledgements}
Research reported in this publication was supported by Cornell University, the Army Research Office under Award Number W911NF-17-1-0247, and the National Institute Of General Medical Sciences of the National Institutes of Health under Award Number R01GM143556.
The content of this manuscript is solely the responsibility of the authors and does not necessarily represent the official views of Cornell University, the U.S.\ Army Research Office, or the National Institutes of Health.

\appendix
\section{\label{API}MModel API} 

We show how to build and interact with the model using the example in Section \ref{example}.

\subsection{Define Functions}

The first step is to define the five functions of the calculation process, shown in Lst.~\ref{lst:functions}.
Note that for $F_y$ and $F_z$ calculations, instead of looping all grid positions, we calculate all the forces in the grid points at once using matrix multiplication of $Q$ and $B_y$ matrices and sum the values of all grid points.
The approach is more efficient with libraries optimized for array programming, such as \textit{Numpy}.\citep{Harris2020sep}
Here we only calculate the force on the magnet when it moves in the positive $x$ direction. For the numerical simulation, we approximate the charge distribution $Q$ in a rectangular prism grid defined by the parameter $grid$. 

\begin{lstlisting}[language=Python, frame=tb, label=lst:functions, caption={Define the functions needed for Lorentz force law model calculating the forces on moving magnet from a charge distribution.}]
import math
import numpy as np


def Bzfunc(mu0, Ms, r, grid, xm):
    """Magnetic field in the z direction.

    :param float mu0: vacuum permeability
    :param float Ms: magnet saturation magnetization
    :param float r: magnet radius
    :param np.ndarry grid: generated by np.ogrid, np.mgrid,
        or np.meshgrid with the shape of (3, x, y, z)
    :param int xm: instantaneous magnet position in
        x direction

    :return: Bz at the relative magnetic-charge position
    :rtype: np.ndarray
    """
    X = (grid[0] - xm) / r
    Y = grid[1] / r
    Z = grid[2] / r
    R = np.sqrt(X**2 + Y**2 + Z**2)
    Bz = mu0 * Ms / 3 * (3 * Z**2 / R**5 - 1 / R**3)
    return Bz


def Byfunc(mu0, Ms, r, grid, xm):
    """Magnetic field in the y direction.

    :param float mu0: vacuum permeability
    :param float Ms: magnet saturation magnetization
    :param float r: magnet radius
    :param np.ndarry grid: generated by np.ogrid, np.mgrid,
        or np.meshgrid with the shape of (3, x, y, z)
    :param int xm: instantaneous magnet position in
        x direction

    :return: By at the relative magnetic-charge position
    :rtype: np.ndarray
    """
    X = (grid[0] - xm) / r
    Y = grid[1] / r
    Z = grid[2] / r
    R = np.sqrt(X**2 + Y**2 + Z**2)
    By = mu0 * Ms * (Y * Z / R**5)
    return By


def Vxfunc(f, A, xm):
    """Magnet velocity at position xm.

    :param float f: cantilever frequency
    :param float A: cantilever amplitude
    :param float xm: instantaneous magnet position in the
        x direction

    :return: instantaneous magnet velocity at position xm
    :rtype: float
    """
    omega = 2 * math.pi * f
    vx = omega * math.sqrt(A**2 - xm**2)
    return vx


def Fyfunc(Q, vx, Bz):
    """Force on the magnet in the y direction.

    :param np.ndarray Q: charge distribution
    :param float vx: instantaneous magnet velocity in the
        x direction
    :param np.ndarray Bz: magnetic field in the z direction

    :return: force in the y direction
    :rtype: float
    """
    Fy = np.sum(Q * vx * Bz)
    return Fy


def Fzfunc(Q, vx, By):
    """Force on the magnet in the z direction.

    :param np.ndarray Q: charge distribution
    :param float vx: instantaneous magnet velocity in the
        x direction
    :param np.ndarray By: magnetic field in the y direction

    :return: force in the z direction
    :rtype: float
    """
    Fz = -np.sum(Q * vx * By)
    return Fz
\end{lstlisting}

\begin{figure*}[t]

\centering
\includegraphics[width=0.8\textwidth]{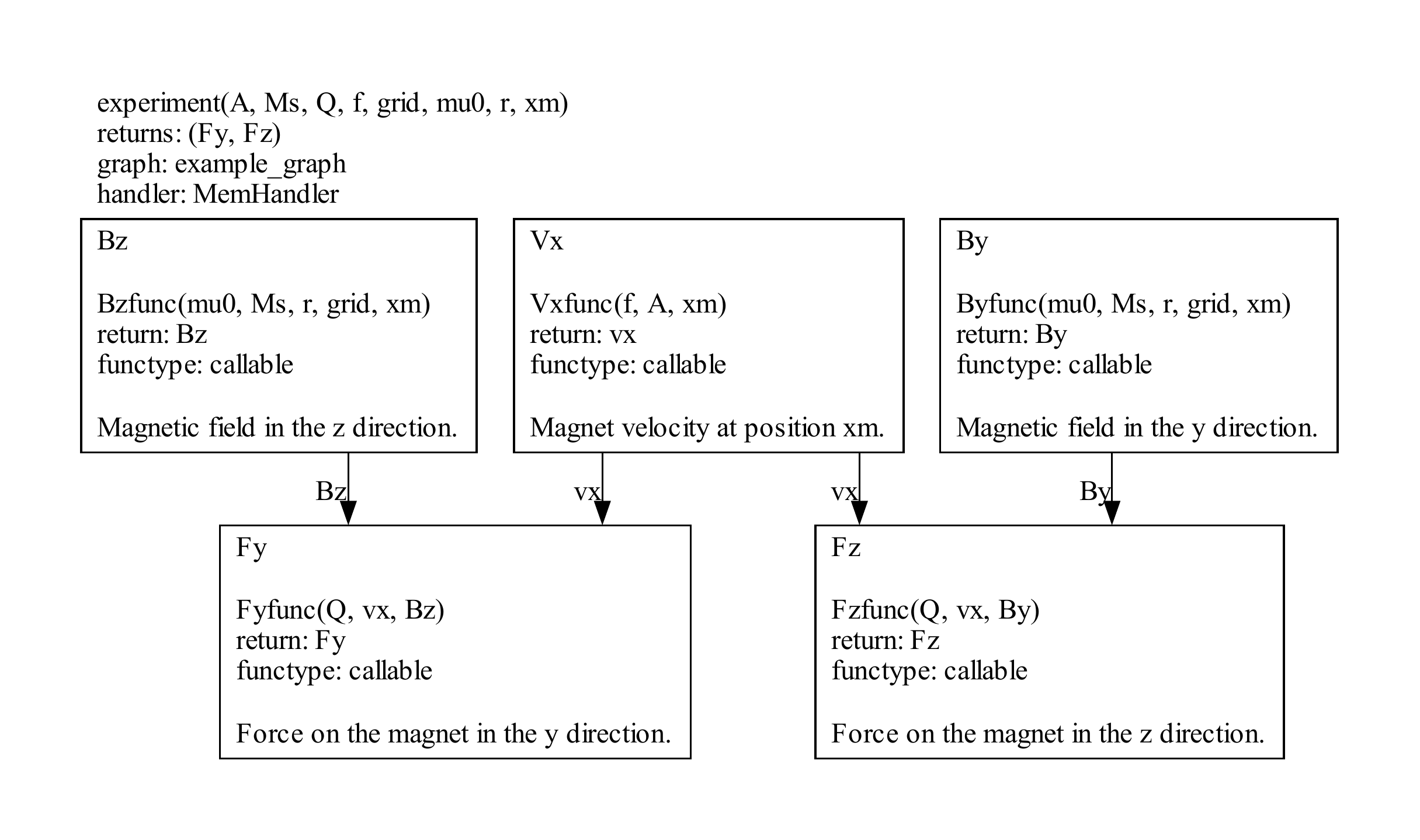}
\caption{Experiment graph model generated by \textit{mmodel}.}

\label{fig:api1}
\end{figure*}

\subsection{Graph and function mapping}

The second step is to define the graph that contains all the function nodes and the edge flow in a DAG.
The graph in \textit{mmodel}  is defined with the \texttt{ModelGraph} class.
The class inherits from \texttt{networkx.DiGraph} of the popular graph library \textit{NetworkX}.\cite{Hagberg2008}
All \textit{NetworkX} operations can be used on \textit{mmodel} graphs.
The nodes and edges can be defined with all available \textit{NetworkX} or \textit{mmodel} syntax for adding notes and edges.
The edges represent the data flow, and the nodes represent the function.
For the graph, the node names and edges are first defined using \texttt{add\_grouped\_edge}, then we map the functions to nodes using \texttt{set\_node\_object}.
Listing \ref{lst:graph} defines the graph G with grouped edges. The functions and output variable name are mapped to the nodes in the \texttt{(node\_name, function, output)} tuple.

\begin{lstlisting}[language=Python, frame=tb, label=lst:graph, caption={Create a \textit{mmodel} graph and map the functions to the corresponding graph nodes, and their output variable names.}]
from mmodel import ModelGraph

G = ModelGraph(name="example_graph")
grouped_edges = [(["Bz", "Vx"], "Fy"), (["By", "Vx"], "Fz")]
G.add_grouped_edges_from(grouped_edges)
node_objects = [
    ("Bz", Bzfunc, "Bz"),
    ("By", Byfunc, "By"),
    ("Vx", Vxfunc, "vx"),
    ("Fy", Fyfunc, "Fy"),
    ("Fz", Fzfunc, "Fz"),
]
G.set_node_objects_from(node_objects)
\end{lstlisting}

\subsection{Model and handler}

The third step is to convert the graph into a model.
The execution method is defined using the handler classes.
The \texttt{Model} class takes graph and handler as input, and the resulting object is a Python callable, and its input arguments are automatically parsed.
As of version 0.5, \textit{mmodel} provides three handlers, \texttt{BasicHandler}, \texttt{MemHandler}, and \texttt{H5Handler}.
The handler information is provided using the handler class, and additional keyword arguments to the handler class can be provided as keyword to the Model class.s
The model can be defined as shown in Lst.~\ref{lst:model}.
The graph of the model is accessible through the \texttt{graph} attribute.

\begin{lstlisting}[language=Python, frame=tb, label=lst:model, caption={Convert the \textit{mmodel} graph to a model by defining the model name and execution method.}]
from mmodel import Model, MemHandler

experiment = Model("experiment", G, handler=MemHandler)
\end{lstlisting}

\begin{table*}[t]
\caption{\label{table:api-execute}Input parameters for the Lorentz force law model, the units are adjusted to match the calculation.}

\begin{ruledtabular}
\begin{tabular}{llll}
 variable & value & unit & description\\ 
 \hline
$A$ & 300 & nm & cantilever amplitude  \\ 
$f$ & 3000 & Hz & cantilever frequency  \\ 
$x_\mathrm{m}$ & 100 & nm & magnet position \\
$r$ & 4000 & nm & magnet radius\\ 
$\mu_0$ & $1.26 \times10^9$ & aN/$\rm{A^2}$ &vacuum permeability \\ 
$M_\mathrm{s}$ & $1.45 \times 10^{-3}$ & A/nm & saturation magnetization \\ 
$Q$ &  $\sim$ 0.5 electron/nm$^3$  & A$\cdot$s & charge distribution \\ 
$grid$ & $2000\times2000\times500$ & nm & grid points \\
voxel &  128000 & nm$^3$ & grid voxel size\\

\end{tabular}
\end{ruledtabular}

\end{table*}

\subsection{Metadata and graph representation}

\textit{mmodel} provides rich metadata and graph plotting capabilities.
The model and graph metadata can be extracted by printing out the string value shown in Lst.~\ref{lst:metadata}. The graph representation uses the dot graph, a standard for representing the graph.
With the \textit{Graphviz} package,\citep{Gansner2000} the users can employ the default graph plotting method in \textit{mmodel} or provide their own.
Both graph and model have the \texttt{draw} method to display the metadata and the graph. The output of \texttt{model.draw()} is shown in Fig.~\ref{fig:api1}.

\begin{lstlisting}[language=Python, frame=tb, label=lst:metadata, caption={Print out \textit{mmodel} graph and model metadata. The graph metadata shows the number of nodes and edges, and the model metadata shows the input, returns, execution method, and modifiers.}]
>>> print(G)
ModelGraph with 5 nodes and 4 edges

>>> print(experiment)
experiment(A, Ms, Q, f, grid, mu0, r, xm)
returns: (Fy, Fz)
graph: example_graph
handler: MemHandler
\end{lstlisting}

\subsection{Model execution}

Here we define representative experimental parameters and execute the model as a function, shown in Lst.~\ref{lst:execute}.
We choose the magnet to be a 250 nm radius cobalt sphere and define a sample grid size of dimension $2000 \: \mathrm{nm}\times 2000  \: \mathrm{nm} \times 500  \: \mathrm{nm}$ in the shape of $(25, 25, 25)$.
The sample is centered at $(0, 0, -600)$ nm, which makes the sample surface 100 nm away from the magnet surface.
The instantaneous magnet position is $(20, 0, 0)$ nm.
The mock sample has a charge distribution of approximately 0.5 electron / nm$^3$, and values are randomly generated.
All the parameters defined are shown in Table.~\ref{table:api-execute}.

\begin{lstlisting}[language=Python, frame=tb, label=lst:execute, caption={Define input parameters and execute \texttt{experiment} model. The resulting forces are in the unit of aN.}]
grid = np.ogrid[-1000:1000:25j, -1000:1000:25j, -350:-850:25j]  # nm
v_voxel = (2000/25)*(2000/25)*(500/25) # nm^3
np.random.seed(0)
Q = np.random.rand(25, 25, 25) * 1.6e-19 * v_voxel  # A s
xm = 100  # nm
A = 300  # nm
f = 3000  # Hz
mu0 = 1.26e9  # aN/A^2
Ms = 1.45e-3  # A/nm
r = 250  # nm

# force in the unit of aN
Fy, Fz = experiment(A, Ms, Q, f, grid, mu0, r, xm)

>>> print("{:.2f} aN".format(Fy), "{:.2f} aN".format(Fz))
6.95 aN 0.15 aN
\end{lstlisting}

\subsection{Graph execution returns}

The model class allows the user to define the return parameters and the return order.
For example, to investigate the intermediate value $v_x$ along with the forces, we simply add $v_x$ to the experiment definition, as shown in Lst.~\ref{lst:intermediate}.

\begin{lstlisting}[language=Python, frame=tb, label=lst:intermediate, caption={Create a model with customized returns and returns order by define the \texttt{returns} parameter.}]
experiment_debug = Model(
    name="experiment_debug", graph=G, handler=MemHandler, returns=["Fy", "Fz", "vx"]
)

>>> print(experiment_debug)
experiment_debug(A, Ms, Q, f, grid, mu0, r, xm)
returns: (Fy, Fz, vx)
graph: example_graph
handler: MemHandler
\end{lstlisting}

\subsection{Modifier}

The modifier is one of the core features of \textit{mmodel}.
The modifiers do not change the original function definition.
Instead, they are Python wrappers that modify the functions after their definition.

For example, the modifier that prints out the node value can be defined as Lst.~\ref{lst:mod}.
The pattern argument specifies the format of the output.
The modifier adds a printing step after the node calculation. 

\begin{lstlisting}[language=Python, frame=tb, label=lst:mod, caption={Define the modifier that prints out the formatted return value of the function after execution.}]
from functools import wraps


def print_result(pattern, end="\n"):
    """Print out function return values.

    :param str pattern: print format
    :param str end: print at the end of the string.
    """

    def stdout_modifier(func):
        @wraps(func)
        def wrapped(**kwargs):
            result = func(**kwargs)
            print(pattern.format(result), end=end)
            return result

        return wrapped
    stdout_modifier.metadata = f"print_result({repr(pattern)}, {repr(end)})"
    return stdout_modifier
\end{lstlisting}

The modifier can decorate any Python function, adding additional steps before or after the execution.
The modifiers can be added to the node or model in a list with their keyword parameters in dictionary form, and multiple modifiers can be applied.
Listing \ref{lst:mod2} adds the modifier to the Vx node and the experiment model, which adds the print steps of formatted $V_x$, $F_y$, and $F_z$ values and their units.

\begin{lstlisting}[language=Python, frame=tb, label=lst:mod2, caption={Apply the \texttt{print\_result} modifier during the function node and model definition.}]
# at the node level
G.set_node_object(
    node="Vx",
    func=Vxfunc,
    output="vx",
    modifiers=[print_result("vx: {:.2e} nm/s", " | ")],
)
# at the model level
experiment = Model(
    "experiment",
    graph=G,
    handler=MemHandler,
    modifiers=[print_result("Fy: {0[0]:.2f} aN | Fz: {0[1]:.2f} aN")],
)

>>> Fy, Fz = experiment(A, Ms, Q, f, grid, mu0, r, xm)
vx: 5.33e+06 nm/s | Fy: 6.95 aN | Fz: 0.15 aN
\end{lstlisting}

\begin{figure*}[t!]

    \centering
    \includegraphics[width=0.7\textwidth]{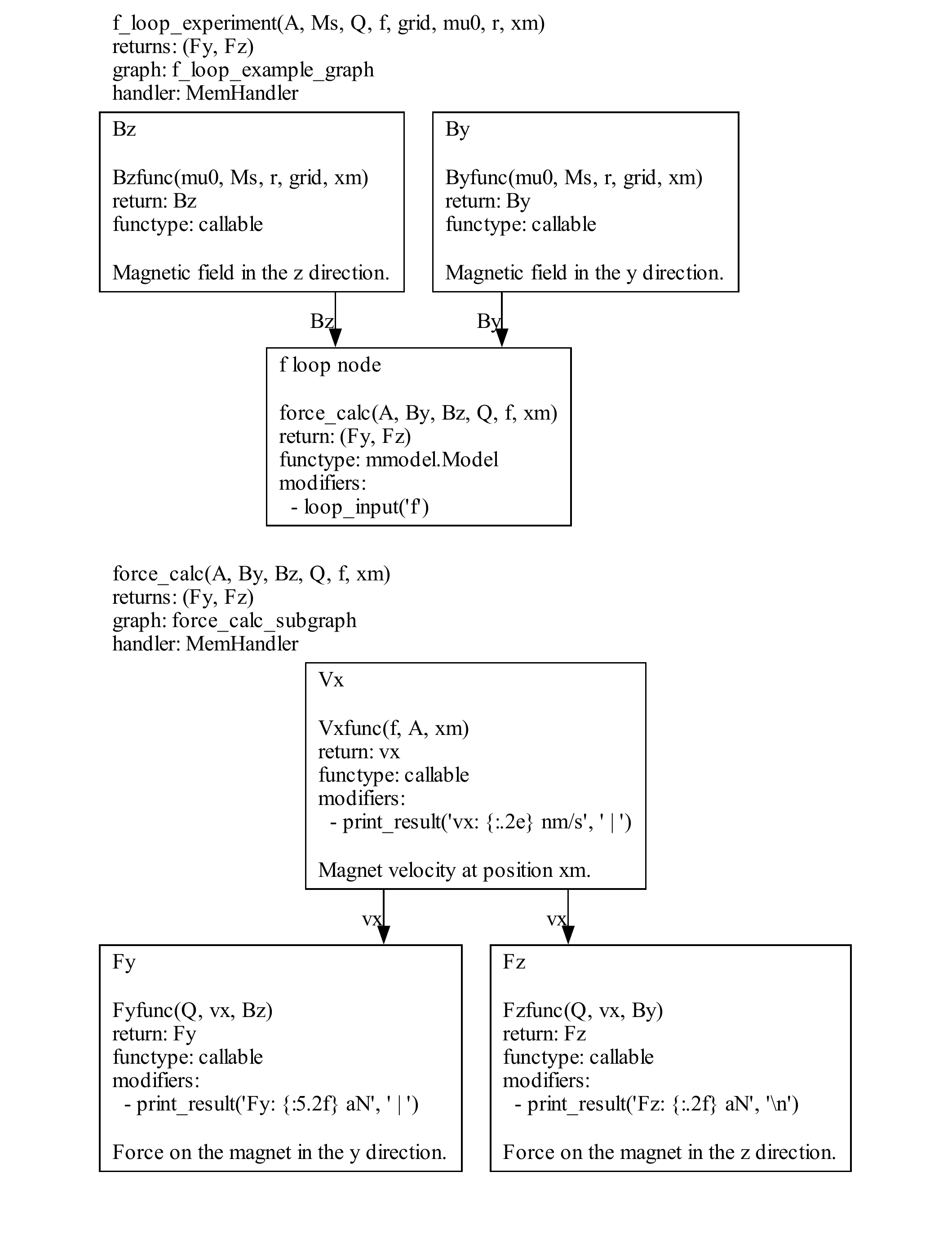}
    \caption{Model graph that loops parameter $f$ generated using the \texttt{loop\_shortcut}. Top: graph of the modified model. Bottom: the graph of the \texttt{f loop node}.}
    
    \label{fig:loop_graph}
\end{figure*}

\subsection{Create loops within the model}
\label{subsec:looping}

In this section, we define a function that can modify the graph of a model and create a loop inside the graph.
We can define a shortcut combining the proper steps of creating the loop discussed in Section~\ref{case:ee}.
The resulting shortcut function, shown in Lst.~\ref{lst:loop-shortcut}, allows the model to take list values of the target parameter.
\textit{mmodel} does not provide the shortcut function for looping since many loop operations can be package specific.

\begin{lstlisting}[language=Python, frame=tb, label=lst:loop-shortcut, caption={Define the shortcut that loops a given parameter within a model. The shortcut modifies the graph and the subnodes.}]
from mmodel.modifier import loop_input


def loop_shortcut(model: Model, loop_param: str, submodel_name: str):
    """Shortcut to create loops of loop_param in the model.

    :param mmodel.Model model: the model to be looped
    :param str loop_param: the loop parameter
    :param str submodel_name: the name of the submodel

    :return: the looped model
    :rtype: mmodel.Model
    """

    G = model.graph
    # determine the subgraph that has the loop parameter as the input
    H = G.subgraph(inputs=[loop_param])
    H.name = f"{submodel_name}_subgraph"

    # create the inner function and substitute the subgraph with the loop_modifier
    loop_func = Model(submodel_name, H, model.handler)

    # determine the subgraph returns
    output = "({})".format(", ".join(H.returns))

    # replace the subgraph, and add the loop_modifier
    loop_graph = G.replace_subgraph(
        H,
        f"{loop_param} loop node",
        loop_func,
        output=output,
        modifiers=[loop_input(loop_param)],
    )
    loop_graph.name = f"{loop_param}_loop_{G.name}"
    model_name = f"{loop_param}_loop_{model.name}"
    return Model(model_name, loop_graph, model.handler)

\end{lstlisting}

The shortcut can be used directly on an existing model, as shown in Lst.~\ref{lst:loop-shortcut-execute}.
The resulting model requires the $f$ input to be iterable, and the results are the iterated ($F_y$, $F_z$) pairs.
The graph is modified to show the proper units of the force nodes.
The resulting model graph and the loop subgraph are shown in Fig.~\ref{fig:loop_graph}.

\begin{lstlisting}[language=Python, frame=tb, label=lst:loop-shortcut-execute, caption={Create a model that loops parameter $f$ based on the Lorentz force model and execute the model with the iterated $f$ value.}]
# Modify the Fy and Fz nodes to output the result.
G.modify_node("Fy", modifiers=[print_result("Fy: {:5.2f} aN", " | ")], inplace=True)
G.modify_node("Fz", modifiers=[print_result("Fz: {:.2f} aN")], inplace=True)
experiment = Model("experiment", G, MemHandler)

# define the loop experiment
>>> loop_experiment = loop_shortcut(experiment, loop_param = "f")
>>> print(loop_experiment)
f_loop_experiment(A, Ms, Q, f, grid, mu0, r, xm)
returns: (Fy, Fz)
graph: f_loop_example_graph
handler: MemHandler

# define frequencies to loop over and execute the experiment
>>> f = np.arange(3000, 8000, 1000) # Hz
>>> loop_result = loop_experiment(A, Ms, Q, f, grid, mu0, r, xm)
vx: 5.33e+06 nm/s | Fy:  6.95 aN | Fz: 0.15 aN
vx: 7.11e+06 nm/s | Fy:  9.26 aN | Fz: 0.20 aN
vx: 8.89e+06 nm/s | Fy: 11.58 aN | Fz: 0.25 aN
vx: 1.07e+07 nm/s | Fy: 13.89 aN | Fz: 0.30 aN
vx: 1.24e+07 nm/s | Fy: 16.21 aN | Fz: 0.35 aN
\end{lstlisting}

Please check out the detailed documentation at \url{https://marohn-group.github.io/mmodel-docs/}.

\newpage

\bibliography{bib/mmodel_bib.bib}

\end{document}